\documentclass
{aa}
\usepackage{fleqn}

\usepackage{url}

\usepackage{graphicx}
\usepackage[figuresright]{rotating}
\usepackage{natbib}
\bibpunct{(}{)}{;}{a}{}{,}
\begin{document}

\title{A Puzzling Paucity of Double Peaked  X-ray Pulsars}
\titlerunning{A Puzzling Paucity...}
\authorrunning{T. Bulik et al.}
\author{T. Bulik \inst{1},
       D. Gondek-Rosi{\'n}ska \inst{2,1},
       A. Santangelo \inst{3},
       T. Mihara \inst{4},
       M. Finger, \inst{5},
       M. Cemeljic\inst{6}}

\institute{$^1$Nicolaus
Copernicus Astronomical Center, Bartycka 18, 00716 Warsaw, Poland\\
          $^2$LUTH, Observatoire de Paris, Place Jules Janssen, F-92 195 Meudon
	  Cedex, France\\
          $^3$IFCAI-CNR, Via Ugo La Malfa 153, 90146 Palermo, Italy\\
	  $^4$RIKEN, 2-1 Hirosawa,  Wako, Saitama 351-0198,  Japan\\
	  $^5$NASA MSFC, Hunstville, AL USA\\
	  $^6$AIP, An der Sternwarte 16,  D-14482  Potsdam, Germany}


\date{Received / Accepted }

\abstract{
Accretion powered pulsars exhibit a	variety of lightcurves.
In this paper we propose to classify the observed lightcurves as 
single or double pulsed.We analyze the lightcurves of 86 accretion powered
pulsars and assign them to these classes. We present three datasets:
first in which the classification can be easily done,  second for which
the classification is more difficult and not certain, and
third for which we were unable to classify the pulsar because of lack
of published data.
We analyze a simple model in which the angles between the magnetic 
and rotation 
axis $\beta$,
and between the rotation axis and the line of sight $\theta$
are random,
and show that it is inconsistent with the data. We also present a model in which
the angle between the magnetic axis and rotation axis is restricted
and  compare it with the data. This leads to an  upper
 limit on 
the angle  $\beta < 40^\circ$. We conclude that there must be
a mechanism that leads to alignment of the magnetic and spin axis
in X-ray pulsars.
}

\maketitle

\section{ The lightcurves of accretion powered
X-ray pulsars}

Accreting neutron stars were discovered nearly 
more than 30 years ago \citep{1967ApJ...148L...1S}, with  Cen X-3
the first one discovered  which showed pulsations
\citep{1971ApJ...167L..67G}. Currently we know nearly one
hundred accreting neutron stars,
and in more than eighty of them pulsations were identified
\citep{2000A&AS..147...25L,2001A&A...368.1021L}.
Accreting neutron stars in binaries exhibit a wide range  of
X-ray light curves. They vary as a function of the photon
energy, and moreover in the transient sources the pulse shapes change 
with the variation of the luminosity.

The pulse period in accreting sources is identified with the
rotation  of a magnetized star. As the
matter from the  companion star falls onto the neutron star
it is channeled onto the magnetic poles by the strong magnetic
field of the neutron star. Thus the polar caps are the places where 
most of the emission takes place.
Several theoretical models of radiation of magnetized accretion 
powered neutron stars have been proposed. 

The magnetized
radiative transfer is solved using a difference scheme
\citep{1985ApJ...298..147M,1992ApJ...395..564B}, or using  Monte
Carlo scheme \citep{1990ApJ...363..670L,1998ApJ...505..688I},
for a review see  \cite{1992QB843.N4M47....}. 
The radiation from each polar cap
produces  one pulse in the light curve.
Depending on the emission cap physics and the strength
of the magnetic field each pulse may have some additional structure.
 When during the rotation of
an accreting  pulsar we see two polar caps the lightcurve should
exhibit two distinct pulses (peaks), and if only one cap is seen then the
lightcurve is single peaked. Each pulsar can be classified 
as single or double peaked depending on the location of the observer and
the inclination between the rotation and the magnetic axes.
In this paper  we  classify the observed pulsars into two classes:
first  the single
pulsed light curves as these for which the geometry is such that
only one cap is seen, and the double pulsed light curves as
those for which two caps are seen. 
At low X-ray energies this
distinction may  not be clear, however in the high energy band  we
clearly distinguish single or double pulses, and there are no
triple (or multiple) pulsed lightcurves, see e.g. 
\citet{1989PASJ...41....1N,1997ApJS..113..367B}. By high energy band we mean
at least above $10$\,keV, or in the pulsars where
cyclotron lines were found, above the cyclotron lines.

The paper is organized as follows: in section 2 we summarize the observations,
 in section 3 we analyze
the expected shapes  and
expected fraction of single and double peaked lightcurves, in
section 4 we discuss the results. We summarize the conclusions in section 5.

\section{Pulsar observations}

In this section we  review the existing observations
of X-ray pulsars and attempt to classify their lightcurves
according to the number of pulses in their lightcurves.
We summarize our classification in Tables 1, 2, and 3.
For each pulsar we have searched the literature for 
X-ray observations, and verified if the pulsar was observed in 
the band above 10\,keV. The pulsars with such observations are listed in
Table~\ref{dobre}. There are 47 entries in the Table out of which
21 have been classified as double and 26 as single.
Additionally we have listed 25 pulsars in Table~\ref{kiepskie} for which
no lightcurve above $10$\,keV was available. There
 are 3 pulsars with likely double light curves and 22 with single
 in Table~\ref{kiepskie}.
 We see that in each table there are more single pulsed than double
pulsed objects. For completion we also list 14 pulsars in Table~\ref{nieznane}
for which
no information on the type of the lightcurve was available.
We define three sets of data corresponding to the three
Tables: dataset ${\cal S}_1$  with the data of Table~\ref{dobre},
dataset ${\cal S}_2$ with the combined data of Tables~\ref{dobre} and~\ref{kiepskie},
and dataset ${\cal S}_3$ with all the data, and
the entries of Table~\ref{nieznane} are
assumed to be double peaked.

\begin{table*}
\caption{X-ray pulsars in group one, i.e. these that are easily classifiable
as single an double pulsed. Part 1}
\begin{tabular}{lllp{4.5cm}p{6cm}}
\hline
\hline
System         &  $P_{spin}$&  Type  &   Comments     &  References \\
\hline
SAX J1808.4-3658& 0.002494  &  Single     &                       & \citet{1998Natur.394..344W}\\
EXO 1024.0-5732& 0.061      &  Double     &       &\citet{1989ApJ...338..338C}\\
GRO J1744-28   & 0.467      &  Single     & The Bursting Pulsar & \citet{1996Natur.381..291F} \\
SMC X-1        & 0.717      &  Double     &                     & \citet{1993ApJ...410..328L}\\
Her X-1        & 1.24       &  Double     &  2nd pulse weak      &\citet{1972ApJ...174L.143T,1997ApJS..113..367B}\\
SMC X-2        & 2.37       &  Single     &   &\citet{2001ApJ...548L..41C}\\
4U 0115+63     & 3.61       &  Double     &       &  \citet{1992ApJ...389..676T,1997ApJS..113..367B}\\
V0332+53       & 4.37       &  Single     &       & \citet{1990PASJ...42..295M}\\
GRO J1750-27   & 4.45       &  Single      &       & \citet{1997ApJ...488..831S}\\
Cen X-3        & 4.82       &  Double     & 2nd pulse weak  &\citet{1971ApJ...167L..67G,1997ApJS..113..367B,2000ApJ...530..429B}\\
1E 2259+586    & 6.98       &  Double     &      &  \citet{1992PASJ...44....9I} \\
4U 1626-67     & 7.66       &  Single     &   &    \citet{1972ApJ...178..281G,1997ApJ...474..414C} \\
GS 0834-430    & 12.3        & Double     &                    &\citet{1992PASJ...44..641A,1997ApJ...479..388W}\\
LMC X-4        & 13.4       &  Single     &        &\citet{1972ApJ...174L.143T,1996ApJ...467..811W}\\
J00521-7319    & 15.3       &  Double    &        & \citet{2001ApJ...560..378F}\\
XTE J1946+274  & 15.8       &  Double &            & \citet{2001AnA...370..529P}\\
2S 1417-624    & 17.6       &  Double    &        & \citet{1997ApJS..113..367B}\\
GRO J1948+32   & 18.7       &  Single    &        &  \citet{1995ApJ...446..826C} \\
RX J0117.6-7330& 22.0       &  Double   &   &  \citet{1999ApJ...518L..99M}\\
GS 1843+00     & 29.5       &  Double    &         &  \citet{1990ApJ...356L..47K}\\
J0112.2-7317   & 31.03      &  Single &              & \citet{1998IAUC.7048....2W}\\
RX J0812.4-3114& 31.9       &  Single    & accretion flow eclipse & \citet{1999MNRAS.306...95R}\\
OAO 1657-415   & 37.7       &  Single    &   & \citet{1978Natur.275..296P,1997ApJS..113..367B}\\
EXO 2030+375   & 41.7       &  Double    & broad peak with two maxima & \citet{1999ApJ...512..313S,1999MNRAS.302..700R}\\
1SAX J0054.9-7226 &  59     &  Single    & & \citet{1998AnA...338L..59S}\\
Cep X-4        & 66.2       &  Double    & & \citet{1991ApJ...366L..19K} \\
XTE J1906+09   & 89.2       &  Single    & double pk below 20keV & \citet{1998ApJ...502L.129M}\\
GRO J1008-57   & 93.5       &  Single    &                   & \citet{1997ApJS..113..367B}  \\
GS 1843-024    & 94.8       &  Single   &     &  \citet{1990PASJ...42L..59K}\\
4U 0728-25     & 103.2      &  Double    &   &\citet{1997ApJ...489L..83C}\\
A0535+26       & 105        &  Double    & accretion flow eclipse    &  \citet{1997ApJS..113..367B}  \\
Sct X-1        &  111       &  Double    &   broad peak two maxima    &  \citet{1991ApJ...370L..77K}\\
GX 1+4         &  120       &  Single    & accretion flow eclipse&   \citet{1997ApJS..113..367B}  \\
GRO J2058+42   &  198       &  Single    &       &\citet{1997ApJS..113..367B}  \\
Vela X-1       &  283       &  Double    &     & \citet{1995ApJ...444..405B,1997ApJS..113..367B}\\
\end{tabular}
\label{dobre}
\end{table*}

\begin{table*}
\caption{X-ray pulsars in group one, i.e. these that are easily classifiable
as single an double pulsed. Part 2}
\begin{tabular}{lllp{4.5cm}p{6cm}}
\hline
\hline
System         &  $P_{spin}$&  Type  &   Comments     &  References \\
\hline
4U 1145-619    &  292       &  Single    &     & \citet{1997asxo.proc..109S,1997ApJS..113..367B}\\
1E 1145.1-6141 &  297       &  Single    &    & \citet{1992SvAL...18..228G} \\
J0103.2-7209   & 345.2      &  single    &    &\citet{2000ApJ...531L.131I}\\
SAX J2103.5+4545& 358.6     &  Single    &   &  \citet{1998AnA...337L..25H}\\
XTE J1855-026 & 361.1       &  Single    &   &  \citet{1999ApJ...517..956C}\\
A 1118-616     &  405       &  Single    &                      &  \citet{1994AnA...289..784C} \\
GPS 1722-363   &  414       &  Single    &        & \citet{1989PASJ...41..473T}\\
4U 1907+09     &  438       &  Double    & 2nd pulse weak &    \citet{1998ApJ...496..386I} \\
4U 1538-52     &  530       &  Double    & &  \citet{1997ApJS..113..367B}  \\
GX 301-2       &  681       &  Double    &  &\citet{1997ApJ...479..933K} \\
X Per          &  835       &  Single    &  &  \citet{1989ApJ...346..469R,1996ApJ...472..341R}\\
RX J0146.9+6121& 1413       &  Single    &     &  \citet{2000AnA...354..567M} \\
\end{tabular}
\label{dobrea}
\end{table*}

\begin{table*}[t]
\caption{X-ray pulsars in group two, i.e.  with no pulse shape above 10\,keV or were
difficult to classify.}
\begin{tabular}{lllp{5cm}p{6cm}}
\hline
\hline
System         &  $P_{spin}$&  Type  &   Comments     &  References \\
\hline
SAX J0635+0533 & 0.0338     &  Single     &   below 10 keV  &\citet{2000ApJ...528L..25C}\\
RX J0059.2-7138& 2.76       &  Single     &   ROSAT obs      & \citet{1994ApJ...427L..25H} \\
RX J0502.9-6626& 4.06       &  Single     &   ROSAT obs  & \citet{1995PASP..107..450S}\\
RX J1838.4-0301& 5.45       &  Single     &   ROSAT obs   &  \citet{1994AnA...286L..47S} \\
1E 1048-593    & 6.44       &  Single     &   Einstein obs  & \citet{1986ApJ...305..814S} \\

AX J1845.0-0300& 7          &  Single     & ASCA obs  & \citet{1998ApJ...503..843T}\\
4U 0142+614    & 8.69       &  Double     & obs below 10 keV & \citet{1994ApJ...433L..25I} \\
2E 0050.1-7247 & 8.88        &  Single    &  ROSAT obs & \citet{1995IAUC.6277....1I} \\
2S 1553-54     & 9.26        &  Double    & SAS-3 obs below 11keV & \citet{1983ApJ...274..765K}\\
AX J0049-732      & 9.31       &   Single     &  ASCA obs   & \citet{2000PASJ...52L..63U}\\
1RXS 170849.0-400910 & 11.0  &  Single     & ASCA obs; could be an AnXP & \citet{1997PASJ...49L..25S}\\
RX J0648.1-4419& 13.2        &  Single    & ROSAT obs  &\citet{1997ApJ...474L..53I}\\
RX J0529.8-6556& 69.5       &   Single    & ROSAT obs & \citet{1997AnA...318..490H}\\
AX J0049-729   & 74.8       &   Single     & ASCA obs & \citet{1999PASJ...51..547Y}\\
SAX J0544.1-710& 96.1       &  Single    & obs below 10keV & \citet{1998AnA...337..772C}\\
1SAX J1324.4-6200&  170       &  Single    & SAX obs below 20keV  & \citet{1998AnA...339L..41A}\\
RX J0440.9+4431&  202.5     &  Single    & obs below 20 keV & \citet{1999MNRAS.306..100R}\\
AX J1749.2-2725&  220       &  Single    & ASCA obs& \citet{1998ApJ...508..854T}\\
GX 304-1       &  272       &  Single    &  SAS-3 obs  &   \citet{1977ApJ...216L..15M} \\
 AX J0058-7203 &  280       &  Double    & ASCA obs& \citet{1999PASJ...51L..21T}\\

AX J0051-733   &  323       &  Single    & ASCA obs & \citet{1999PASJ...51L..15I}\\
1SAX J1452.8-5949 &  437       &  Single    & obs MECS below 10keV & \citet{1999AnA...351L..33O}\\
AX J170006-4157& 714.5      &  Single & ASCA obs & \citet{1999ApJ...523L..65T}\\
RX J1037.5-564 &  860       &  Single    &  pulse shape below 20 keV  & \citet{1999MNRAS.306..100R}\\
2S 0114+650    &10008     &   Single    & ASM obs below 12 keV&\citet{1999ApJ...511..876C,2000ApJ...536..450H}\\
\end{tabular}
\label{kiepskie}
\end{table*}

\begin{table*}[t]
\caption{X-ray pulsar in group three, for which no reliable pulse shapes were found.}
\begin{tabular}{lllp{5cm}p{6cm}}
\hline
\hline
System         &  $P_{spin}$&  Type  &   Comments     &  References \\
\hline
A0538-67       & 0.069      &   ?        & no lc published        &   \citet{1997ApJS..113..367B}        \\
J0501.1-7211   & 3.343      &  ?         & no lc published & \citet{1998IAUC.7028....1Y}\\
RX J0720.4-3125& 8.38       &  ?         &  isolated ns accreting from ISM?&   \citet{1996IAUC.6445....2H}\\
EXO 053109-6609.2& 13.7     &  ?         & EXOSAT detection & \citet{1995IAUC.6184....2D}\\
J0053.8-7226   & 46.6       &  ?        & no lc  published    & \citet{1998IAUC.6803....1C}\\
X 1839-04      &  81.1      &   ?     & no lc published   &   \citet{1990Natur.343..148K}\\
J0051-722      & 91.1       & ?  &   & \citet{2001AnA...368.1021L}\\
AX 1820.5-1434 &  152       &  ?     &  weak pulsations  &  \citet{1998ApJ...495..435K}\\
J0054-720       & 169.3     & ?        & no lc  published & \citet{1998IAUC.6814....1L}\\
3U 1239-599   &  191        &  ? &  Ariel obs, no lc published & \citet{1977MNRAS.180P..21H}\\
XTE J1858+034 &  221        &  ? & no lc published  &  \citet{1998IAUC.6826....2R}\\
4U 2206 + 54  & 392         &     ?  & no lc published & \citet{1992ApJ...401..678S}\\
1WGA J1958.2+3232& 730      &   ? & only ROSAT lc &\citet{1998MNRAS.298..502I}\\
AX J0049.5-7323&  755       &   ?    & weak pulsations & \citet{2000PASJ...52L..73Y}\\
\end{tabular}
\label{nieznane}
\end{table*}

\section{ The expected pulsars light curves}

\begin{figure}[t]
\begin{center}
\includegraphics[width=0.9\columnwidth]{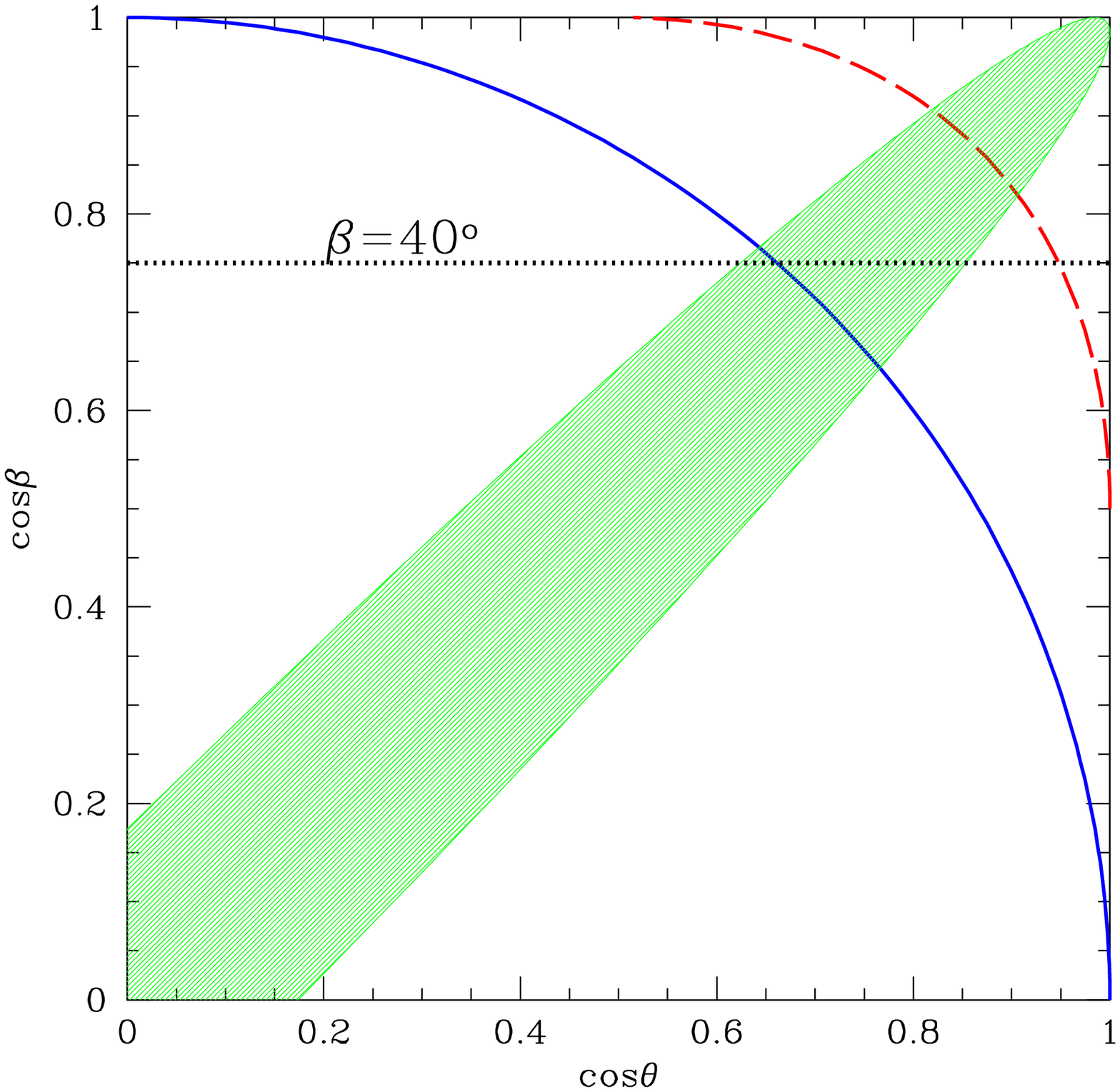}
\end{center}

\caption{Parameter space for viewing accreting X-ray pulsars.
The region below the solid line corresponds to
visibility of two polar caps with gravitational light bending
neglected. The region below the dashed line  
corresponds to visibility of two caps with the gravitational
light deflection $\Delta_g=30^\circ$. The  shaded
region is the part of the parameter space in which
pulses would be detectable if radiation is strongly beamed.
 The horizontal  dotted line corresponds to the angle between
the rotation and magnetic axes $\beta=40^\circ$.} 
\label{pspace} 
\end{figure}

Let us assume that an accreting neutron star has a pure centered
dipole field. In such case the geometry of accretion powered
pulsars is characterized by two angles: $\theta$ - the angle
between the rotation axis and direction to the observer, and
$\beta$ the angle between the magnetic and rotation axes. The
parameter space when only one cap is seen is determined by the
condition: 
\begin{equation}
 \theta < 90-\beta-\Delta_g \, ,
\end{equation} 
 where $\Delta_g$ is
the gravitational light deflection of a ray  emitted parallel to
the surface of the neutron star. Here we assume that the 
magnetic beaming is not very strong and that there are X-ray
photons emitted parallel to the surface of the neutron star.
Ignoring the gravitational light bending effects, i.e. in the
classical Newtonian case, $\Delta_g =0^\circ$,
the parameter space  for seeing just one cap is shown as the
region below the solid line in
Figure~\ref{pspace}.
 Thus, if the angles $\beta$ and $\theta$ are chosen randomly the
fraction of single pulsed objects should be $1-\pi/4 \approx 0.214$. 
The region  for the single  peaked light curves would be
smaller if the radiation were strongly beamed. However, in this
case the double peaked light curves would exhibit narrow pulses,
contrary to what is observed - see e.g. Figure~7 in
\citet{1997ApJS..113..367B}.

\begin{figure}[t]
\begin{center}
\includegraphics[width=0.9\columnwidth]{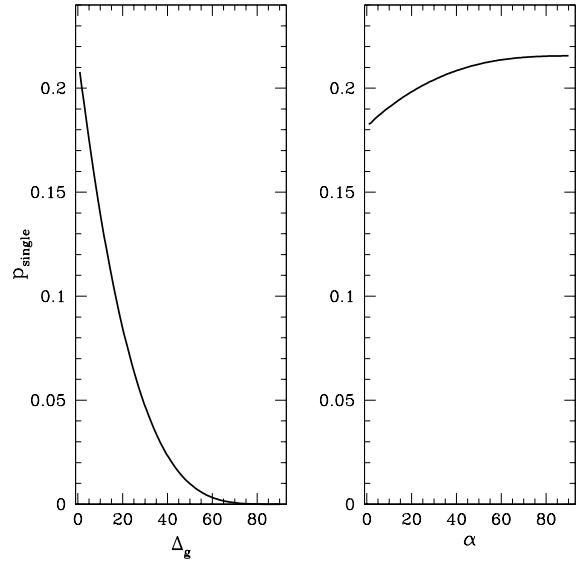}
\end{center}

\caption{Left panel: the probability of seeing only one polar cap
as a function of the maximal gravitational light deflection. The non-relativistic
case corresponds to $\Delta_g=0$. Right panel; the probability of seeing
only one cap as a function of the beaming angle $\alpha$.}
\label{delal}
\end{figure}

For a set of data, like pulsars, with a well defined 
property, i.e. either single or double pulsed, and a 
given probability $p$ that each pulsar will belong to one or another class
the probability of observing $n$ single  peaked lightcurves and 
$m$ double peaked in an observation of randomly chosen 
$n+m$ objects is given by the binomial distribution:
\begin{equation}
P_{n,m} = { (n+m)!\over n! m! } p^n (1-p)^m \,, 
\end{equation}
and the probability of obtaining  $M$ or less double peaked 
objects in a sample containing $N$ source is a  sum 
\begin{equation}
{\cal P}_{N,M} = \sum_{m=1}^M P_{N-m,m}\,.
\label{probequ}
\end{equation}
We can test the consistency of the datasets defined above 
with the simplest  model: random inclination of the magnetic axis to
the rotation axis $\beta$,
random direction of the line of sight $\theta$, and
no gravitational light deflection.
For the dataset  ${\cal S}_1$ we obtain the probability $2.66\times 10^{-7}$
that there are 21 or less double pulsars in a set of 47 observed objects.
Applying this test to the dataset ${\cal S}_2$ we obtained the probability
of only $9.33\times 10^{-17}$, and for the dataset ${\cal S}_3$
this probability is $1.76\times 10^{-12}$. Thus we see
that the data are inconsistent with this simple model.
This is due to the fact that the parameter space for viewing
single pulsed objects is smaller than the double pulsed objects.
Including the effects of the gravitational light deflection 
only reduces the probability of seeing one pulsed objects and makes 
the values of the above probabilities drop even further.
We present the probability of seeing one polar cap as a function of
$\Delta_g$ in the left panel of Figure~\ref{delal}.
Another possibility to consider is that the pulsar beam is not wide
but strongly collimated, and the beam has an opening angle 
of $\alpha$.  In this case for a given $\beta$ there
is only a small range of angles $\beta-\alpha <\theta<\beta+\alpha $ 
for which the pulsar is 
visible at all. The restriction on the parameter space for small $\alpha=10^\circ$
is shown in Figure~\ref{pspace} as the shaded region. 
The probability of seeing only one accretion cap as a function of the 
angle $\alpha$ is shown in the right panel of Figure~\ref{delal}.
Another possibility to consider is to allow for non
aligned dipole, i.e. to assume that the two polar caps
are not antipodal but can be placed anywhere on
the stellar surface. A straightforward calculation neglecting the
gravitational light bending effects shows
that the probability of seeing just one polar cap drops
in this case to 17\%.
Thus the probabilities computed above in the Newtonian case
are conservative
and inclusion of the  gravitational  light bending and beaming
make the result even more significant.

\begin{figure}
\begin{center}
\includegraphics[width=0.9\columnwidth]{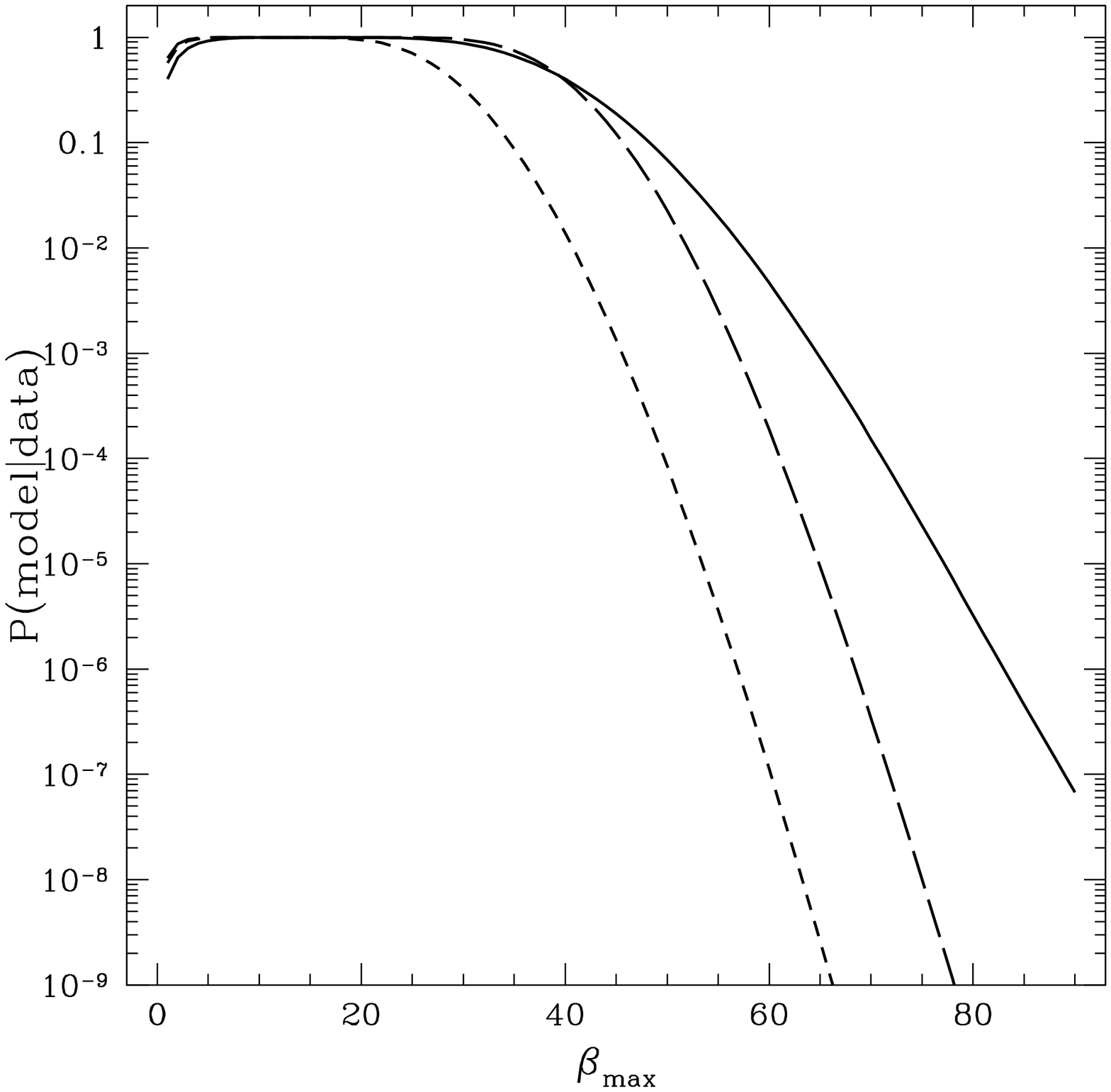}
\end{center}
\caption{The probability of observing the data
in models parameterized by the the maximum angle
between the magnetic and rotation axis $\beta_{max}$.
The solid line corresponds to the data set ${\cal S}_1$,
and the short-dashed line corresponds to the data set ${\cal S}_2$.
The long-dashed line corresponds to the data set ${\cal S}_3$. }
\label{signifnr}

\end{figure}

The incompatibility of the simplest model with the data 
prompts us to consider a more general model. We assume first that
the distribution of the inclinations of the magnetic
axes to the rotation axis is restricted, i.e. the angles
$\beta$ are limited to the range between $0$ and $\beta_{\rm max}$.
This corresponds to restricting the parameter space to the 
region above a horizontal line at $\cos\beta_{\rm max}$ in Figure~\ref{pspace}.
For each value of $\beta_{\rm max}$ we can compute 
the probability of observing the given set of data using
equation~(\ref{probequ}). We plot this probability 
as a function of $\beta_{\rm max}$ in Figure~\ref{signifnr}.
The standard model corresponds to the case $\beta_{\rm max}= 90^\circ$
and the probability of seeing the data 
rises to nearly unity for the angles $\beta_{\rm max} < 40^\circ$
for the datasets ${\cal S}_1$ and ${\cal S}_3$, and 
$\beta_{\rm max} < 30^\circ$ for the 
dataset ${\cal S}_2$. The probabilities of seeing one cap 
increases with decreasing $\beta_{\rm max}$, and therefore
for its intermediate values the probability of seeing the data 
is rising to nearly unity. For the very small values  of
$\beta_{\rm max}$ one expects to see  nearly all pulsars 
as single pulsed. Since each dataset contains some double 
pulsed objects the probability of seeing the data drops
when $\beta_{\rm max}\rightarrow 0$.

\begin{figure*}
\begin{center}
\includegraphics[width=0.3\textwidth]{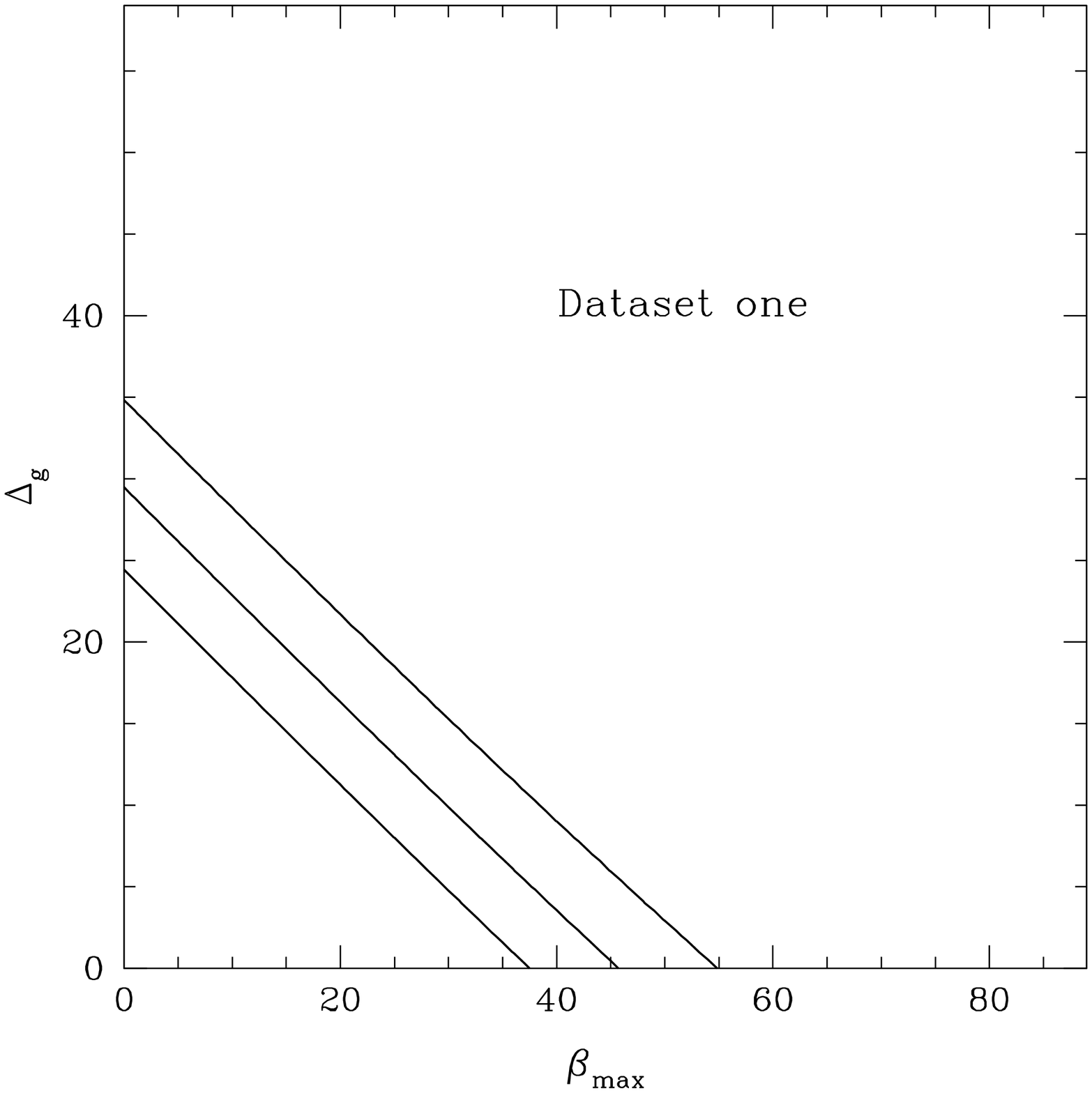}
\includegraphics[width=0.3\textwidth]{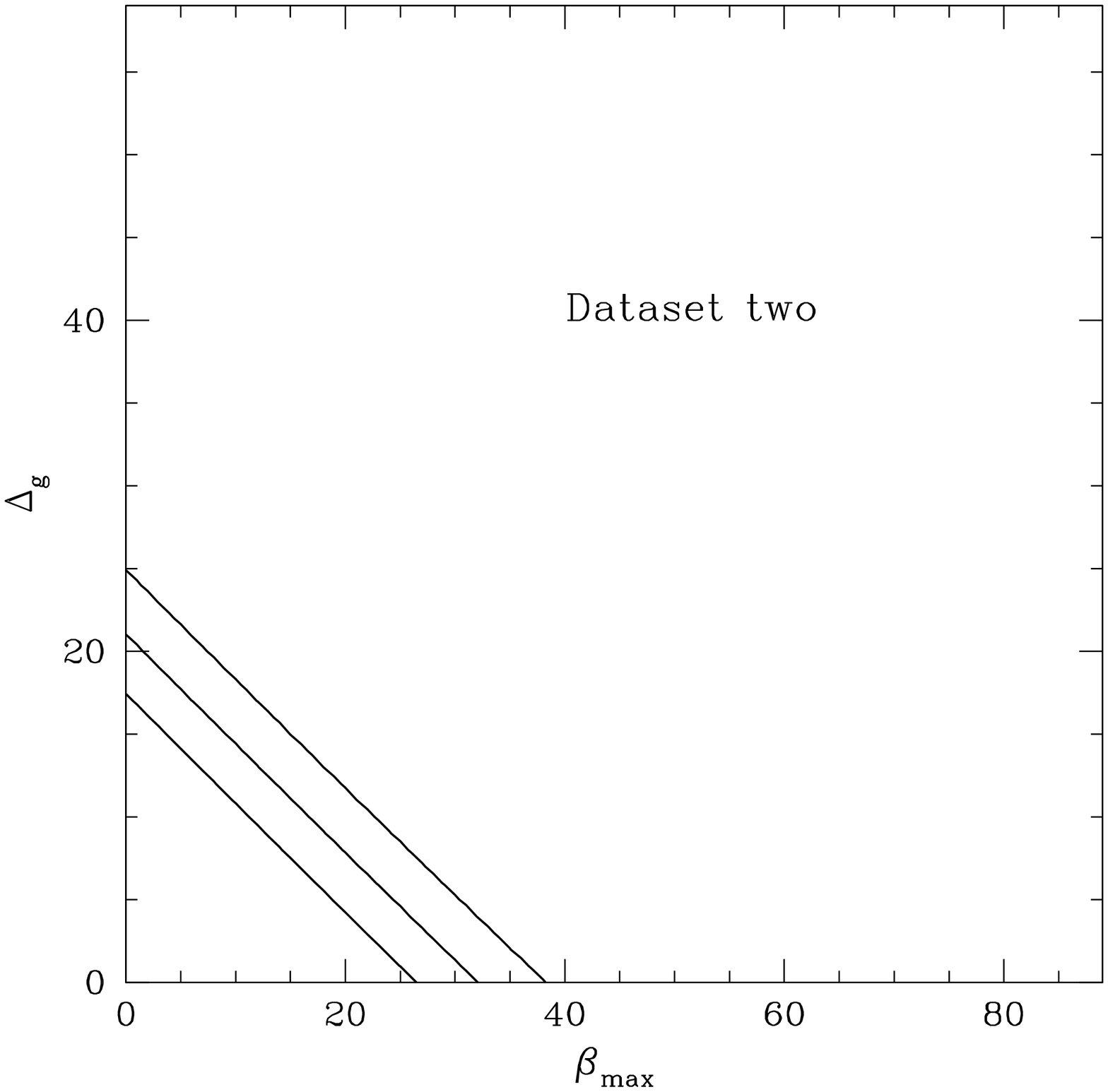}
\includegraphics[width=0.3\textwidth]{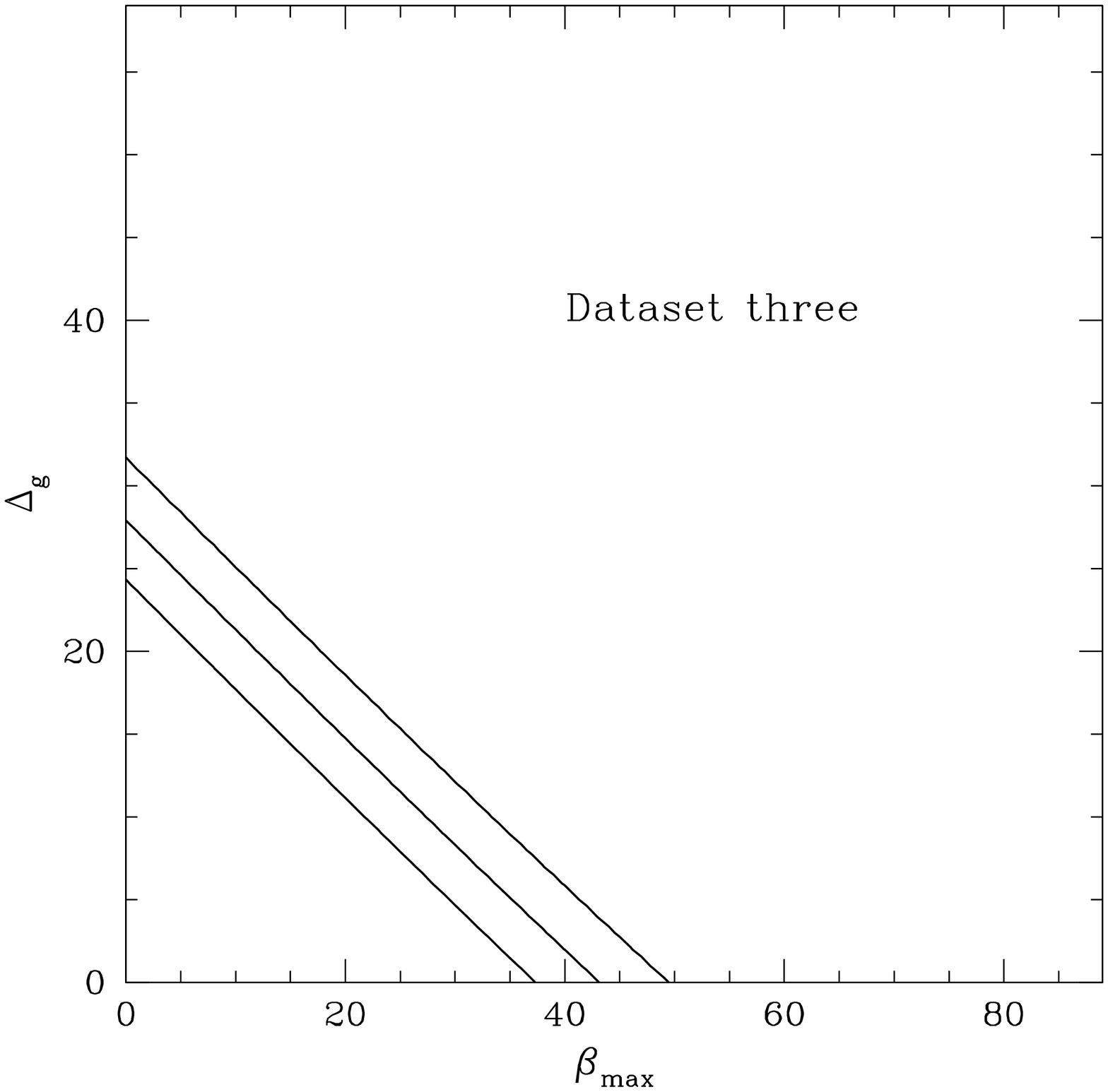}
\end{center}
\caption{Each panel presents the region containing
the models that are consistent with the data at the level
of 60\%, 90\%, and 99\%. The three panels correspond
to the three datasets.}
\label{panls}
\end{figure*}

A more general model taking into
account the general relativistic effects includes two parameters: the
maximal angle between the magnetic and rotation axis $\beta_{\rm max}$
and the  maximum gravitational light deflection angle
$\Delta_g$. For such a model we compute the probability
of seeing a pulsar with only one polar cap as a function of these two angles.
In Figure~\ref{panls} we present the
regions containing  the models that are consistent with
the data with the probability  higher than 40\%, 10\%, and 1\%.
The non relativistic results correspond to
the line $\Delta_g=0$ in Figure~\ref{panls}.
Since the probability of seeing only one polar cap decreases
with increasing  $\Delta_g$, we also set a limit
on its maximum value. The region corresponding to the
1\% agreement limit for the dataset ${\cal S}_1$ can
be expressed as $\Delta_g + 0.63 \beta_{\rm max}< 33^\circ $.
Thus if there were no gravitational light deflection
effects  we find that the inclination angle of the magnetic
axis to the rotation axis should be smaller than $\approx 50^\circ$.
For the case with a moderate gravitational light bending
$\Delta_g= 20^\circ$ the magnetic axis inclination must
be smaller than $\beta_{\rm max}< 20^\circ $. We also
see that the amount of gravitational light bending
cannot exceed $33^\circ$ at the 99\% confidence level.
The limits are even stronger when we take into account
dataset ${\cal S}_2$, however for the set ${\cal S}_3$
the constraints are similar to these obtained
for the dataset ${\cal S}_1$.

\begin{figure}[t]
\begin{center}
\includegraphics[width=0.9\columnwidth]{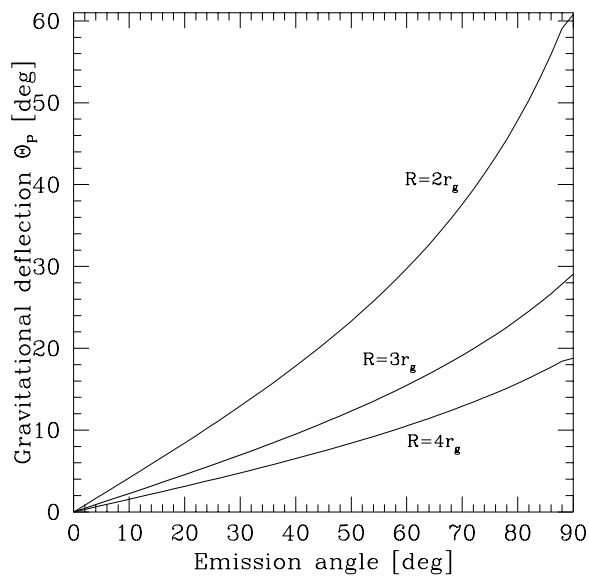}
\end{center}
\caption{The value of the gravitational light bending as a
function of the emission angle, measured from the radial direction
for photons emitted at different initial radii. The
maximum gravitational light deflection is the value of deflection
for the initial angle of $90^\circ$, i.e. parallel to the
surface.}
\label{gravbend}
\end{figure}

\begin{figure}[t]
\begin{center}
\includegraphics[width=0.9\columnwidth]{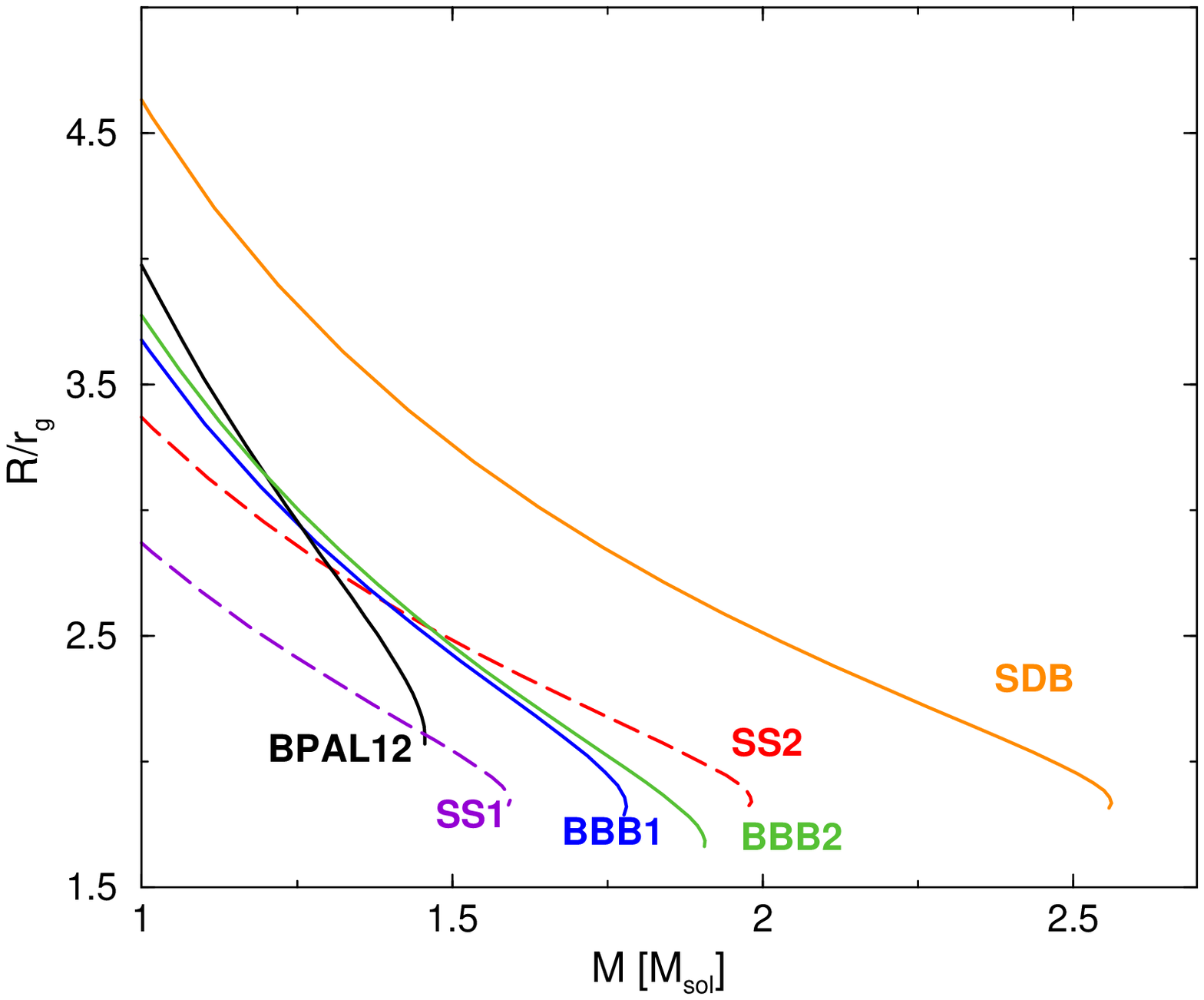}
\end{center}
\caption{The ratio of the stellar radius to the gravitational radius
as a function of mass for several neutron star equations of
state. The labels are BPAL12 - \citet{1997PhR...280....1P},
SBD - \citet{1993ApJ...416..267S}, BB1 and BB2 -
\citet{1997AnA...328..274B}, while SS1 and SS2 correspond to the
MIT Bag model of strange quark matter \citep{Witten84} with two different
densities at zero pressure.
}
\label{mrg}
\end{figure}

\section{Discussion}

To discuss the limit on $\Delta_g$ obtained above 
we will consider two cases: (i) emission from a polar
cap, and (ii) emission from the accretion column. In the first 
case the upper limit on the $\Delta_g$ obtained above 
can be interpreted as a limit on the compactness
of the star. We plot the amount of gravitational
light deflection as a function of the emission angle
for a few values of the ratio of the neutron star radius to the
gravitational radius $r/r_g$ in Figure~\ref{gravbend}.
The maximal gravitational light deflection of about 
$30^\circ$ corresponds to the case with ratio $r/r_g \approx 3$ for the
photons emitted parallel to the surface.
The limit on $\Delta_g$ can be understood as the upper
limit on the compactness of the neutron stars.
In Figure~\ref{mrg} we plot the ratio $r/r_g$ as a function of
the stellar gravitational mass for a few representative equations
of state.  The ratio $r/r_g$ decreases with increasing mass.
The upper limit on the ratio $r/r_g$ corresponds
for a given equation of state to the lower limit 
on the mass of the accreting neutron stars. Masses
of neutron stars in some accreting objects have been
estimated for from the observations of the binary motion, like
e.g. in the case of 4U1538-52
\citep{1983A&A...122...79P,1992MNRAS.256..631R} and they all seem
to be consistent with the fiducial $1.4\,M_\odot$ with
some uncertainty. The general consensus seems to be
that  the masses of all neutron stars are consistent with 
the mean value of $\approx 1.4$\,M$_\odot$, and this is also
the value found for the  binary pulsars
 \citep{1999ApJ...512..288T},
however in LMXBs this range may be much wider 
and reach above $2\,M_\odot$ \cite{2000AnA...361..153B}.
If the neutron stars have masses around $1.4\,M_\odot$ and
the beams are wide
then our results point to the
very stiff equations of  state like e.g. SDB \citep{1993ApJ...416..267S}.

A second possibility is that we see the emission 
from the accretion column  with
the typical properties of the fan beam and that this emission is responsible
for the observed flux at angles with respect to the normal
above 90 degrees. Since we limit the width of the beam
to $\Delta_g <33^\circ$. In this case our limit for the non-relativistic
case means that the emission from the column extends no
higher than to $\approx 0.2 r$. If the effects of the gravitational
light bending around a neutron star are included the
maximum height allowed is even smaller . Thus our results
argue for the polar emission and against significant
emission from the accretion column.

We have obtained a
a limit on the possible inclination angles of the magnetic axis
with respect to the rotation axis $\beta_{max}< 50^\circ$.
A hint of such distribution was already shown by
\citet{1991MNRAS.251..203L}. He analyzed the lightcurves
of 20 pulsars using a simple model with the beam dependence on the
emission angle as $\cos^2\phi$ and $\cos^4\phi$ where
$\phi$ is the direction to the normal of the surface.
For each pulsar he found the angles  $\theta$ and $\beta$
however he fits were degenerate in these angles, so that
it was not clear a priori which angle is which. The resulting
distribution   was inconsistent with the assumption that both
angles are random, and \citet{1991MNRAS.251..203L} concluded
that the mean angle between the magnetic and rotation axis
is as small $15^\circ$. \citet{1995ApJ...444..405B}
modeled  lightcurves of 4U~1538-52 and Vela~X-1 observed by Ginga,
and obtained the inclination angles below $50^\circ$ for their
fits with best significance. \citet{1996ApJ...467..794K} analyzed the
pulse profiles of Cen~X-3 and reconstructed the geometry and
beam pattern for this object. They found that the most likely
inclination angles of the magnetic axis to the rotation axis
are $\sim 20^\circ$.

\citet{1998AcA....48...65C} have noticed that the light curves
of  some accretion powered pulsars exhibit sudden dips. Such
dips can be interpreted as accretion flow eclipses, i.e. be
caused by the absorption of the X-ray flux when the accretion
column  passes the line of sight. Such phenomena have been noted
for at least three objects: A0535-262
\citep{1998AcA....48...65C}, GX1+4 \citep{2000ApJ...529..447G},
and LS992 \citep{1999MNRAS.306...95R}.

If such eclipses happen when the accretion column passes through
the line of sight then they should take place when $|\theta
-\beta| < \gamma$, where $\gamma$ is the angular width of the
accretion column. The region corresponding to $\gamma=10^\circ$
is shaded in Figure~\ref{pspace}. We see that such
eclipses should frequently take place in double pulsed light
curves, if the magnetic axis is randomly oriented with respect
to the rotation axis. Moreover within the single pulsed light
curves  a significant fraction, - approximately $20$-$30$\%
should exhibit such dips, and this fraction increases as the
amount of gravitational  light bending increases.

Yet the only three cases when such dips have been found are in
single  pulsed light curves. The lightcurve of A0535-262 seems
to be double peaked, however \citet{1998AcA....48...65C} pointed
out that it can be interpreted as a single pulse divided by the
accretion  flow eclipse. This lack of double peaked pulsars with
accretion flow eclipses (dips) may mean  that the region  of
parameter space for double peaked light curves with dips is
either empty or scarcely populated!  Again we are drawn to the
conclusion  that the magnetic axis needs to be aligned with
the rotation axis. If the magnetic axis inclination is restricted
to the region $\beta_{max}< 50^\circ$, than one expects that the
accretion flow eclipses to be observed  only in the
one pulsed lightcurves, as is the case. While this by itself is
not a significant result, the identification of accretion flow
eclipses is consistent with the conclusion that
the range of allowed values of $\beta$ is limited.

One should mention here that there
may possibly be a  selection bias against seeing double pulsed objects.
In the case of nearly perpendicular rotators the two
pulses will be similar. It is therefore possible that
the lack of double pulsed objects is due to misidentification of the
pulse period which should be two times longer as reported.
The fraction of such misidentified objects depends on the
exact shape of the beam. For wide uncollimated beams
the two pulses look similar for a range of angles $\beta$
as large as the width of the beams. However, our limit
is $\beta_{max}< 50^\circ$, so the beams would have to be very
wide. Additionally, the beam shape
would have to be very fine tuned in order to compensate the viewing
geometry effects over such large range of angles.

The sample of accreting pulsars presented here is dominated by transient
Be/X-ray sytems. One should note that for the disk fed persistent
sources (SMC X-1, SMC X-2, Cen X-3, LMC X-4, RX J0648.1-4419) there
are two double peaked objects out of five sources, still
below the expected $\approx 79\%$, but this is not a significant discrepancy.
In the case of wind fed persistent sources
(Vela X-1, 1E 1145.1-6141, 4U1907+09, 4U1583-53, GX 301-2, 2S 0114+650)
there are four double peaked out of six objects.
However, because of the small number of these sources it is
impossible to obtain any significant conclusion just from
statistical analysis. However, as already mentioned above
detailed modeling of two wind fed pulsars (4U 1538-52 and Vela X-1)
and one disk fed (Cen X-3) resulted in low values of $\beta$.

\section{Conclusions}

We analyzed the  shapes of accretion powered pulsar light
curves, looking for the expected fraction of single and double
peaked light curves, as well as these exhibiting  accretion flow
eclipses. We find that the expected number  of single peaked
lightcurves is small; certainly less than about one in five
lightcurves should be single peaked. On the other hand the
observations show that quite a large fraction of
accreting pulsars have single peaked lightcurves in hard X-rays.
 In
order to explain that we are drawn to conclude that  the
magnetic axes in accretion powered pulsars tend to be aligned
with the rotation axis. Such alignment decreases the expected
number of double peaked lightcurves, as seen in
Figure~\ref{pspace}. We have compared the models
parametrized by the maximal inclination of the magnetic to the
rotation axis $\beta_{max}$ and the value of the
maximum gravitational light deflection $\Delta_g$ with the
data. We obtain upper limits at the 
99\% confidence level $\beta_{max}< 50^\circ$ 
and $\Delta_g < 33^\circ$.

Our results are based on the classification of the neutron star lightcurves
as single or double pulsed. In order to avoid 
possible misclassification we have attempted to use 
the data above $10$~keV, or above the cyclotron line wherever possible.
Moreover, we have attempted to classify as many pulsars
as double when we were in doubt.
In order to rid of the possible systematic effects
we have repeated the analysis for the three datasets, 
and in the third dataset we included all the unclassifiable
pulsars as the double peaked objects to be conservative.
The deficit of double pulsed objects is significant for
every dataset that we analyzed. Moreover we
note that the paucity of the double peaked
pulsars is so strong that our results still hold even if there are still a few
misidentifications in our analysis.
We have also shown that if the pulsar beam is strongly collimated 
the deficit of double peaked objects becomes more significant.

\begin{acknowledgements}
This research has been supported by the KBN grants 2P03D02117
and PBZ-KBN-054/P03/2001, and
the EU Programme "Improving the Human Research Potential and the
Socio-Economic Knowledge Base" (Research Training Network Contract
HPRN-CT-2000-00137).
TB is grateful for the hospitality of the Observatoire
de Paris in Meudon and for the support of the CNRS/PAN Jumelage programme.
\end{acknowledgements}


\end{document}